\documentclass[grl]{agutex}

  \usepackage{graphicx}
\DeclareGraphicsExtensions{.gif,.eps,.png,.tif,.pdf}
\authorrunninghead{ARDHUIN ET AL.}

\titlerunninghead{OCEAN SWELL DISSIPATION}

\authoraddr{Fabrice Ardhuin,
Service Hydrographique et Oc\'{e}anographique de la Marine, 29609
Brest, France. (ardhuin@shom.fr)}

\authoraddr{Bertrand Chapron,
Laboratoire d'Oc\'{e}anographie Spatiale, Ifremer, Centre de
Brest, 29280 Plouzan\'{e}, France. (bertrand.chapron@ifremer.fr)}

\authoraddr{Fabrice Collard, Collecte Localisation Satellites,
division Radar, 29280 Plouzan\'{e}, France. (Dr.fab@cls.fr)}

\begin{document}

%
%

\title{Observation of swell dissipation across oceans}
%

%
%


\author{Fabrice Ardhuin}
\affil{Service Hydrographique et Oc\'{e}anographique de la Marine,
Brest, France}

\author{Bertrand Chapron}
\affil{Ifremer, Laboratoire d'Oc\'{e}anographie Spatiale, Centre
de Brest, Plouzan\'{e}, France}

\author{Fabrice Collard}
\affil{Collecte Localisation Satellites, division Radar,
Plouzan\'{e}, France}



%
%
%

%
%


\begin{abstract}
Global observations of ocean swell, from satellite Synthetic
Aperture Radar data, are used to estimate the dissipation of swell
energy for a number of storms. Swells can be very persistent with
energy e-folding scales exceeding 20,000 km. For increasing swell
steepness this scale shrinks systematically, down to 2800 km for the steepest
observed swells, revealing a
significant loss of swell energy. This value corresponds to a normalized energy decay in time $\beta=4.2 \times 10^{-6}$~s~$^{-1}$. Many processes may be
responsible for this dissipation. The increase of dissipation rate
in dissipation with swell steepness is interpreted as a laminar to turbulent
transition of the boundary layer, with a threshold Reynolds
number of the order of 100,000. These observations of swell evolution open 
the way for
more accurate wave forecasting models, and provides a constraint
on swell-induced air-sea fluxes of momentum and energy.
\end{abstract}

%
%

%

\begin{article}

%
%

\section{Introduction}
Swells are surface waves that outrun their generating wind, and
radiate across ocean basins. At distances of 2000~km and more from
their source, these waves closely follow principles of geometrical
optics, with a constant wave period along geodesics, when
following a wave packet at the group
speed \citep[e.g.][]{Snodgrass&al.1966,Collard&al.2009}. These
geodesics are great circles along the Earth surface, with minor
deviations due to ocean currents.

Because swells are observed to propagate over long distances,
their energy should be conserved or weakly
dissipated \citep{Snodgrass&al.1966}, but  little quantitative
information is available on this topic. As a result, swell heights are relatively poorly predicted \citep[e.g.][]{Rogers2002,Rascle&al.2008}. Numerical wave
models that neither account specifically for swell dissipation,
nor assimilate wave measurements, invariably overestimate
significant wave heights ($H_s$) in the tropics.
  Typical biases in such models reach 45 cm or 25\% of the mean observed wave height in the East
  Pacific \citep{Rascle&al.2008}. Further, modelled peak periods along the North American west coast exceed those
  measured by open ocean buoys, on average by 0.8~s \citep{Rascle&al.2008}, indicating an excess of long period swell energy.
    Theories proposed so far for nonlinear wave evolution or air-sea interactions  \citep[e.g.][]{Watson1986,Tolman&Chalikov1996}, require order-of-magnitude
     empirical correction in order to produce realistic wave heights  \citep[e.g.][]{Tolman2002d}.
     Swell evolution over large scales is thus not understood.

Swells are also observed to modify air-sea interactions
\citep{Grachev&Fairall2001}, and swell energy has been suggested
as a possible source of ocean mixing \citep{Babanin2006}. A
quantitative knowledge of the swell energy budget is thus needed
both for marine weather forecasting and Earth system modelling.

The only experiment that followed swell evolution at oceanic
scales was carried out in 1963. Using in situ measurements, a very
uncertain but moderate dissipation of wave energy was found
\citep{Snodgrass&al.1966}. The difficulties of this type of analysis
are twofold. First, very few storms produce swells that line up
with any measurement array, and second, large errors are
introduced by having to account for island sheltering. Qualitative
investigations by \cite{Holt&al.1998} and
\cite{Heimbach&Hasselmann2000} demonstrated that a space-borne
synthetic aperture radar (SAR) could be used to track swells
across the ocean, using the coherent persistence of swells
along their propagation tracks. Building on these early studies,
\cite{Collard&al.2009} demonstrated that SAR-derived swell heights
can provide estimates of the dissipation rate. Here we make a
systematic and quantitative analysis of four years of global SAR
measurements, using level 2 wave spectra \citep{Chapron&al.2001b}
from the European Space Agency's (ESA) ENVISAT satellite. The
swell analysis method is briefly reviewed in section 2. The
resulting estimates of swell dissipation rates are interpreted in
section 3, and conclusions follow in section 4.

\section{Swell tracking and dissipation estimates}
 Our analysis uses a two step method. Firstly, using
SAR-measured wave periods and directions at different times and
locations, we follow great circle trajectories backwards at the
theoretical group velocity. The location and date of a swell source
is defined as the spatial and temporal center of the convergence
area and time of the trajectories. We define  the spherical
distance $\alpha$ from this storm center ($\alpha = X/R$ where $X$
is the distance along the surface on a great circle, and $R$ is
the Earth radius).

Secondly, we chose a wave period $T$ and,  starting from the
source at time $t=0$ and an angle $\theta_0$, we follow imaginary wave packets
along the great circle at the group speed $Cg=gT/(4 \pi)$.  SAR
data are retained if they are acquired within 3~hours and 100~km
from the theoretical position of our imaginary wave packet, and if a swell
partition is found with peak wavelength and direction within 50~m and
20$^\circ$ of their expected values. 
This set of SAR observations constitutes one swell track.
We repeat this procedure by first varying $\theta_0$.
 Tracks with
neighboring values of $\theta_0$ are merged in relatively narrow
direction bands (5 to 10$^\circ$ wide) in order to increase
the number of observations along a track. This ensemble of tracks is the basic
dataset used in our analysis. Such track ensembles are produced for different
storms and different wave periods. Because the SAR sampling
must match the natural swell propagation, ten storms only produced 22 track
ensembles with enough SAR data that satisfies our selection criteria in the period 2003 to 2007. 
These criteria are wind speeds less than 9~m~s$^{-1}$, 
swell heights larger than 0.5~m, and the observations should span more than 
3000 km along the great circle, in order to produce a stable estimate of the 
swell spatial decay rate $\mu$. 

In the absence of dissipation (i.e. $\mu=0$), \cite{Collard&al.2009} demonstrated
that, in any chosen direction $\theta_0$ and at the spherical
distance $\alpha$ and time $t$ corresponding to a propagation at a
chosen group speed $C_g$, the swell energy $E_s$ decreases
asymptotically as $1/[\alpha \sin(\alpha)]$. The $\sin(\alpha)$
factor arises from the initial spatial expansion of the energy
front, with a narrowing of the directional spectrum. The $\alpha$
factor is due to the dispersive spreading of the energy packet,
because $C_g$ is proportional to $T$, associated to a narrowing of
the the frequency spectrum. \cite{Collard&al.2009} also showed
that for realistic wave conditions $E_s$ should be within 20\% of
the asymptotic values for distances $\alpha R$ larger than 4000~km
from the storm center, where $R$ is the Earth radius. In our estimation of $\mu$,  data within 4000 km of the originating storm are ignored to make sure
that the remaining data are in the far field of the storm. 

This 4000~km value was estimated for a storm of radius $r=1000$~km. This applies to any storm provided that all the energy for the wave period $T$ is confined within this radius at $t=0$, with no generation of such long swells for $t>0$. Fast moving and long-lived storms may lead to larger values of $r$ and, following \cite{Collard&al.2009}, deviations from the asymptote larger than 20\%. An extreme situation would be a steady storm moving along the great circle at the speed $C_g$, that would generate
a constant swell energy $E_s$ as a function of $\alpha$. No such situation was found in the storms analyzed below. 

\begin{figure}
\noindent\includegraphics[width=20pc]{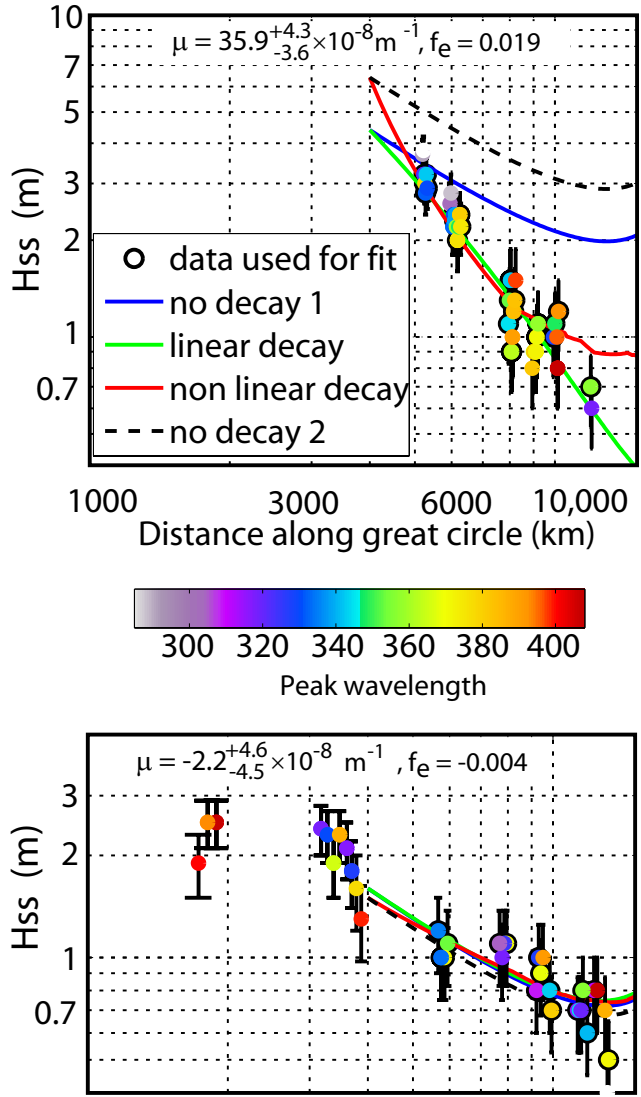}
\caption{(a) Observed swell wave height as a function of distance,
and theoretical decays with fitted constant coefficients using no
dissipation, linear ($\mu$ constant) or non-linear ($f_e$ constant) dissipation, for the 15~s waves
generated by a very strong (top) North Pacific storm on 12
February 2007 (auxiliary table 1: swell number 18) and a weaker
(bottom) southern ocean storm on 12 August 2007 (auxiliary table
1: swell number 19). Circled dots are the observations used in the
fitting procedure. Error bars show one standard deviation of the
expected error on each SAR measurement \citep{Collard&al.2009}.}
\end{figure}

In each track ensemble, all swells have close initial directions
$\theta_0$, and the wave field is only a function of $\alpha$. We
define the spatial evolution rate
\begin{equation}
\mu=-\frac{\mathrm{d} \left(\alpha \sin \alpha E_s\right)/\mathrm{d}\alpha}{R \left(\alpha \sin \alpha E_s\right)}.
\end{equation}
Positive values of $\mu$ correspond to losses of wave energy
(Figure 1.a). Negative, but not significant, values are occasionally found (figure 1.b). 

For each track ensemble we take a reference distance $\alpha_0=\pi/5$ which corresponds to 4000~km. $\mu$ is estimated by finding the
pair  $\left(\widehat{E}_s(\alpha_0),\mu\right)$, that minimizes the
mean square difference between  observed swell energies $E_s(\alpha_i)$ with $i$ ranging from 1 to $N$,
and the theoretical constant linear decay, 
\begin{equation}
\widehat{E}_s(\alpha_i) \alpha_i \sin \alpha_i=\widehat{E}_s(\alpha_0) \alpha_0 \sin \alpha_0 {\mathrm e}^{-R \mu (\alpha_i-\alpha_0)}.
\end{equation}
Because we only have two parameters $\mu$ and $\widehat{E}_s(\alpha_0)$ to adjust, the minimization is performed by a complete search of the parameter space.

\cite{Collard&al.2009} estimated that
the  SAR-derived swell heights $H_{ss}=4\sqrt{E_s}$ are
gamma-distributed about a true value $H_{ss}-b_H$. The bias is well approximated by
\begin{equation}
b_H=0.11 + 0.1 H_{ss}-0.1 \max\{0,U_{10SAR}-7\}\label{Hserr1}
\end{equation}
with $H_{ss}$ in meters and the wind speed $U_{10}$ in
m~s$^{-1}$. A realistic model of the standard deviation of the measurement error is
\begin{equation}
\sigma_H=0.10\mathrm{m}+\min\left\{0.25 H_{ss},0.8 \mathrm{m} \right\}\label{Hserr2}.
\end{equation}
Using this error model,
we generated 400 synthetic data sets by perturbing independently each measured swell
wave height, in order to obtain a confidence interval for $\mu$. For each swell case, the values of $\mu$ and $H_{ss}(\alpha_0)$ reported below
are the medians of the 400 calculated values. 

For all our swell data, $\mu$ ranges from -0.6 to $3.7\times
10^{-7}$ m$^{-1}$ (Figure 2.a), comparable to 2.0$\times
10^{-7}$~m$^{-1}$ previously reported for large amplitude swells
with a 13~s period\citep{Snodgrass&al.1966}. Clarifying earlier
observations by \cite{Darbyshire1958} and
\cite{Snodgrass&al.1966}, our analysis unambiguously proves that
swell dissipation increases with the wave steepness. We recall
that, in the absence of dissipation, a maximum 20\% deviation of $E_s$ relative to the asymptote is expected due to the storm
shape. This deviation is equal to the one produced by a real  5.0$\times
10^{-8}$~m$^{-1}$ dissipation over 4000~km. Thus a comparable error on the estimation of $\mu$ is expected when, as we do here, the storm shape is not taken into account \citep{Collard&al.2009}.

\section{Interpretation of swell dissipation}
At present there is no consensus on the  plausible causes of the
loss of swell energy \citep{WISE2007}. Interaction with oceanic
turbulence is expected to be relatively small
\citep{Ardhuin&Jenkins2006}. Observed modifications and reversals
of the wind stress over swells \citep{Grachev&Fairall2001} suggest
that some swell momentum is lost to the atmosphere. The
wave-induced modulations of air-sea stresses yield a flux of energy from
the waves to the wind, due to the correlations of pressure and
velocity normal to the sea surface, and the correlations of shear
stress and tangential velocity. An upward flux of
momentum, readily observed over steep laboratory waves, can thus result in a
wave-driven wind \citep{Harris1966}.  If these modulations  are linearized
\citep[e.g.][]{Kudryavtsev&Makin2004}, the swell
dissipation rate becomes linear in terms of the wave energy, with a
proportionality constant that typically depends on the wind, but which does
increase with the swell steepness, as we observe here.
\begin{figure}
\noindent\includegraphics[width=20pc]{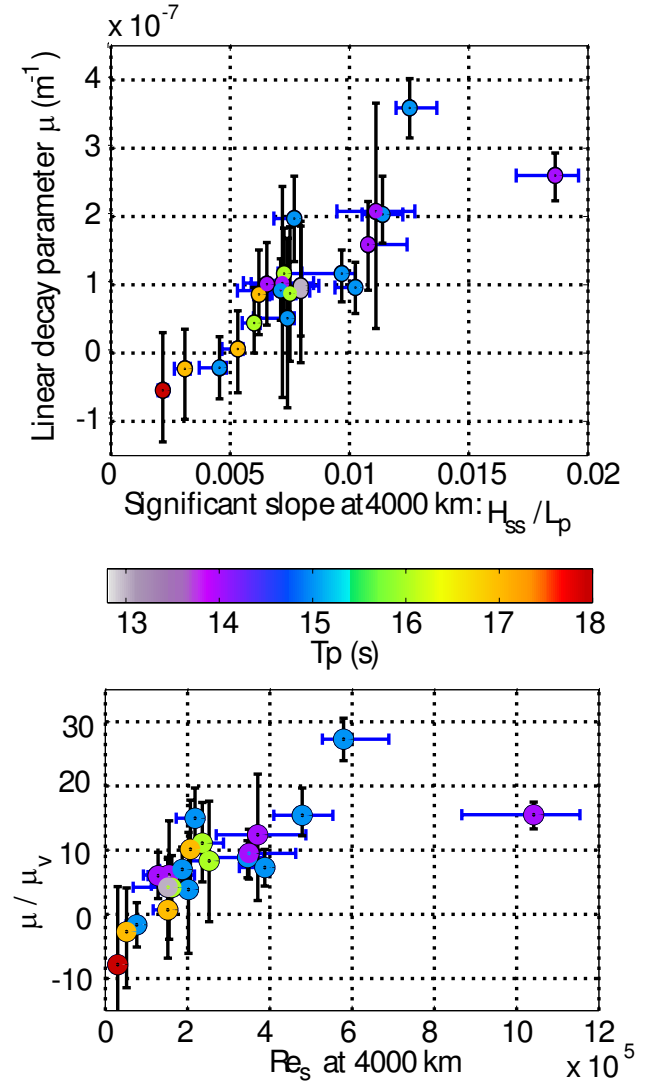}
\caption{Swell dissipation for 22 events (see auxiliary material for details). (a) Estimated linear
attenuation coefficient  as a function of the initial significant
slope, ratio of the swell significant wave height and peak 
wavelength, $s=4 H_s/L$, taken 4000 km from the storm centre, for
a variety of peak swell periods (colors). (b) Attenuation
coefficient   normalized by the viscous attenuation  $\mu_\nu$
(eq. \ref{munu}), as a function of the significant swell Reynolds number Re$_s$
determined from significant velocity and displacement amplitudes at
4000 km from the storm.}
\end{figure}

Our observations show no clear trend with wind magnitude $U_{10}$
and wind-wave angle $\theta_w$: the swell age $C/U_{10}$ or $C/(U_{10} \cos \theta_w)$ averaged over the swell 
track gives little correlation with $\mu$, even when wheigthed with the swell energy. We thus take a novel approach, and interpret our data by
neglecting the effect of the wind, considering only the shear
stress modulations induced by swell orbital velocities. Little
data are available for air flows over swells, but boundary layers
over fixed surfaces are well known, and should have similar
properties if their significant orbital amplitudes of velocity and
displacement are doubled \citep{Collard&al.2009}.
The dissipation then depends on the surface roughness and a
significant Reynolds number, $\mathrm{Re}(\varphi )=
4u_{\mathrm{orb}}(\varphi) a_{\mathrm{orb}}(\varphi)/\nu$, where $u_{\mathrm{orb}}$ and $a_{\mathrm{orb}}$ are the significant amplitudes of the surface orbital velocities and displacements. 

For Re$<10^5$, the flow should be laminar \citep{Jensen&al.1989}.
The strong shear above the surface makes the air viscosity important, with a dissipation coefficient given by \cite{Dore1978} and \cite{Collard&al.2009}
\begin{equation}
\mu_\nu=2 \frac{\rho_a }{\rho_w g C_g }\left(\frac{2 \pi}{T}\right)^{5/2}\sqrt{2 \nu},\label{munu}
\end{equation}
where $L$ is the swell wavelength, $L=g T^2/(2\pi)$ in deep water
with $g$ the acceleration of gravity. At ambient temperature and
pressure, the air viscosity is $\nu=1.4\times
10^{-5}$~m$^2$s$^{-1}$, and $\mu_\nu$ is only a function of $T$.
As $T$ increases from 13 to 19~s, $\mu_\nu$ decreases from
$2.2\times10^{-8}$ to $5.8 \times 10^{-9}$m$^{-1}$.

For larger Reynolds numbers the flow becomes turbulent. The energy rate of decay in time can be written as
 \begin{equation}
\beta=-\frac{\mathrm{d} E_s/\mathrm{d}t}{E_s}=C_g \mu=\frac{\rho_a 4 \pi^2}{\rho_w g T^2} f_e
u_{\mathrm{orb}}
\end{equation}
where $f_e$ is a swell dissipation factor. For a smooth surface, $f_e$ is of the order of 0.002
to 0.008 \citep{Jensen&al.1989}, when assumed equal to the friction factor $f_w$.

Re is difficult to estimate from the SAR data only, because
ENVISAT's ASAR does not resolve the short windsea waves. However,
in deep water we can define the smaller `swell Reynolds number'
$\mathrm{Re_s}$ from $u_{\mathrm{orb,s}}= 2\sqrt{E_s}2\pi/T$ and
$a_{\mathrm{orb,s}}= 2\sqrt{E_s}$. 

Our estimates of $\mu$ exceed $\mu_\nu$ by a factor that ranges
from $O(1)$ to 28 (Figure 2.b), 
quantitatively similar to oscillatory boundary layer over
fixed surfaces with no or little roughness. Namely, dissipation
rates $\mu$ of the order of the viscous value $\mu_\nu$ are found
for Re$_s<5\times 10^4$ when the the flow may be laminar, and we
only find large values of $\mu / \mu_\nu$ when Re$_s
> 5\times 10^4$ over a significant portion of the swell track. For reference, a 6.3 m s$^{-1}$
wind can generate a fully-developed wind-sea with Re$=2\times 10^5$, making the
boundary layer turbulent for any swell amplitude. Using the numerical wave model described in \cite{Ardhuin&al.2009}, one finds that this value of Re$_s$
translates to Re$\simeq 10^5$. That same model also gives values of 
$u_{\mathrm{orb}}$. Fitting a constant $f_e$ for each track ensemble
yields $-0.001 \leq f_e \leq 0.019$, with a median of
0.007, close to what is expected over a smooth surface. This suggests that the roughness of the
waves for this oscillatory motion is very small compared to the orbital amplitude.

A parameterization of swell dissipation, taking $f_e$ constant in the range 0.0035 to 0.007, generally yields accurate wave heights (not shown). The quality of the end result also depends on the other parameterizations for wind input, whitecapping and wave-wave interactions, and requires a rather lengthy discussion \citep[e.g.][]{Ardhuin&al.2008d,Ardhuin&al.2009}.

Beyond this simple model, we expect that winds should modify the boundary layer over swell, with a significant effect for winds larger than 7 m~s$^{-1}$ \citep{Collard&al.2009}. \cite{Kudryavtsev&Makin2002}
considered the wind stress modulations due to short wave roughness
modulated by swells, and found that the preferential breaking of short waves near long wave crests could double the wind-wave coupling coefficient $\mu$ for the long waves.  Yet, their linear model cannot
explain the nonlinear dissipation observed here, because they only
considered lowest order effects. Further investigations should
probably consider both wind and finite amplitude swell effects
to explain the observed variability of $\mu$.


If this dissipation is due to the proposed air-sea friction
mechanism, the associated momentum flux  $\rho_w g E_s/2$ goes to
the atmosphere. If, on the contrary, underwater processes dominate, an energy
flux  $\rho_w g C_g E_s$ may go into ocean turbulence.
Accordingly, these fluxes are small. For 3~m high swells, the
momentum flux is 8\% of the wind stress produced by a 3~m
s$^{-1}$ wind. This momentum flux thus plays a minor role in
observed O(50\%) modifications of the wind stress at low
wind\citep{Drennan&al.1999,Grachev&Fairall2001}. Wind stress modifications are
more likely associated with a nonlinear influence of swell on
turbulence in the atmospheric boundary layer
\citep{Sullivan&al.2008}. 
This effect may arise as a result of the low-level wave-driven wind 
jet \citep{Harris1966} and its effects on the wind profile around the 
critical level for the short wave generation \citep{Hristov&al.2003} . 
Whatever the actual process, the dissipation coefficient $\mu$ is a
key parameter for validating theoretical and numerical models 
\citep{Kudryavtsev&Makin2004,Hanley&Belcher2008}.

\section{Conclusions}
Using high quality data from a space-borne synthetic aperture
radar, ocean swells were systematically
tracked across ocean basins over the years 2003 to 2007. Ten storms 
provided enough data to allow a total of 22
estimations of the swell energy budget for peak periods of 13 to 18~s. 
The dissipation of
small-amplitude swells is not distinguishable from viscous
dissipation, with decay scales larger than 20000~km. On the
contrary, steep swells lose a significant fraction of their
energy, up to 65\% over a distance as short as 2800~km. This
non-linear behavior is consistent with a transition from a laminar
to a turbulent air-side boundary layer. Many other processes may contribute to the observed dissipation, and a full model of the air-sea interface will be needed for further progress. 
The present observations and analysis opens
the way for a better understanding of air-sea fluxes in low wind
conditions, and more accurate hindcasts and forecasts of sea states \citep[see ][ and e.g. the SHOM results in \textit{Bidlot} 2008]{Ardhuin&al.2008d,Ardhuin&al.2009}\nocite{Bidlot2008}. 

Further investigations are necessary to understand the wind stress
modulations and their variations with wind speed, direction, and
swell amplitude. Such an effort is essential for the 
improvement of numerical wave models and their application to remote sensing and the estimation of air-sea fluxes.


%
%
%
%
%
%

%
%
%
%

\begin{acknowledgments}
SAR data was provided by the European Space Agency (ESA). The
swell decay analysis was funded by the French Navy as part of the
EPEL program. This work is a contribution to the ANR-funded
project HEXECO and DGA-funded project ECORS.
\end{acknowledgments}

%
%
%
%
%
%
%
%
%
%


\newcommand{\noopsort}[1]{} \newcommand{\printfirst}[2]{#1}
  \newcommand{\singleletter}[1]{#1} \newcommand{\switchargs}[2]{#2#1}


%
%

\end{article}




%
%
%
%
%
%


\end{document}